\documentclass[aip,pof,amsmath,amssymb]{revtex4-1}
\usepackage{CJK}
\usepackage{amssymb,euscript,latexsym,amsmath}
\usepackage{graphicx}
\usepackage[dvips]{epsfig}
\usepackage{color}
\usepackage{epstopdf}
\usepackage{setspace}
\usepackage{subfigure}
\usepackage{slashbox}
\usepackage{chngpage}
\usepackage{hyperref}

\usepackage{caption2}

\DeclareGraphicsExtensions{.eps,.ps,.pdf}

\begin{document}

\title{Cilia beating patterns are not hydrodynamically optimal}

\author{Hanliang Guo}
\affiliation{Aerospace and Mechanical Engineering, University of Southern California, 854 Downey Way, Los Angeles, CA 90089-1191}
\author{Janna Nawroth}
\affiliation{Wyss Institute for Biologically Inspired Engineering, Harvard University, 3 Blackfan Circle, Boston, MA 02115}
\author{Yang Ding}
\author{Eva Kanso\footnote{Corresponding author: Kanso@usc.edu}}
\affiliation{Aerospace and Mechanical Engineering, University of Southern California, 854 Downey Way, Los Angeles, CA 90089-1191}
\singlespacing

\begin{abstract}
We examine the hydrodynamic performance of two cilia beating patterns reconstructed from experimental data.  In their respective natural systems, the two beating patterns correspond to: (A) pumping-specialized  cilia, and (B)  swimming-specialized cilia. We compare the performance of these two cilia beating patterns as a function of the metachronal coordination in the context of two model systems:  the swimming of a ciliated cylinder and the fluid pumping by a ciliated carpet.  Three performance measures are used for this comparison: (i) average swimming speed/pumping flow rate; (ii) maximum internal moments generated by the cilia; and (iii) swimming/pumping efficiencies. We found that, in both models, pattern (B) outperforms pattern (A) in almost all three measures, including hydrodynamic efficiency. These results challenge the notion that hydrodynamic efficiency dictates the cilia beating kinematics, and suggest that other biological functions and constraints play a role in explaining the wide variety of cilia beating patterns observed in biological systems.
\end{abstract}

\maketitle


\section{Introduction}
\label{sec:intro}

Cilia are slender hair-like structures, typically $5$ to $25$ microns in length, that extend from the surface of eukaryotic cells. Current literature distinguishes various cilia subtypes that differ in morphology and motility \cite{ibanez-tallon2003}. Their internal structure, however, is remarkably conserved across species ranging from protists to humans \cite{vincensini2011}. Motile cilia, the focus of this study, serve a wide range of functions requiring fluid movement, including sensing and transport \cite{bloodgood2010}. In aquatic invertebrate and single-cell species, motile cilia commonly occur along external and internal epithelia, where they are used for locomotion, sensing and feeding \cite{tamm2014}. In mammals, motile cilia serve transport functions such as clearing mucus in the respiratory system, breaking the left-right symmetry during embryonic development, and moving egg cells in the female reproductive system \cite{satir2007}, and increasing evidence hints at their important role in detecting and relaying mechanical and chemical signals \cite{bloodgood2010}. Not surprisingly, cilia defects are associated with a variety of diseases \cite{fliegauf2007}; however, apart from association studies \cite{chilvers2003}, there is limited understanding of how structural and kinematic variation relate to ciliary function. This is partly due to a current trend in biofluid mechanics research to focus on computing the ideal stroke kinematics for a single function, such as optimal fluid transport in cilia, acting both individually \cite{eloy2012} and collectively \cite{osterman2011}. This trend is part of a general disposition towards computing optimal strokes in various microsystems, including Purcell's three-link swimmer \cite{tam2007}, an ideal elastic flagellum \cite{spagnolie2010}, a biflagellated organism \cite{tam2011}, a shape-changing body\cite{avron2004}, a two- and a three-sphere swimmer\cite{alouges2009}, and a squirmer model of a ciliated spherical organism\cite{michelin2010}. Here, we present a study suggesting that such analysis provides valuable insights for the design of biological and artificial micro-propulsion but fails to explain, let alone evaluate, the structural and functional variation arising from multitasking within natural environments.

\begin{figure}[!h]
\centerline{\includegraphics[width=0.5\textwidth]{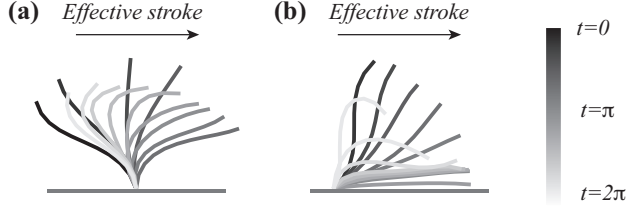}}
  	\caption[]{Beating pattern of an individual cilium. The effective stroke is shown in dark grey and recovery stroke in light grey. (a): rabbit tracheal cilia~\cite{sanderson1981, fulford1986}; (b): protozoan {Opalina}  \cite{sleigh1968}}
	\label{fig:beatingpattern}
\end{figure}

\begin{figure}[t]
\centerline{\includegraphics[width=\textwidth]{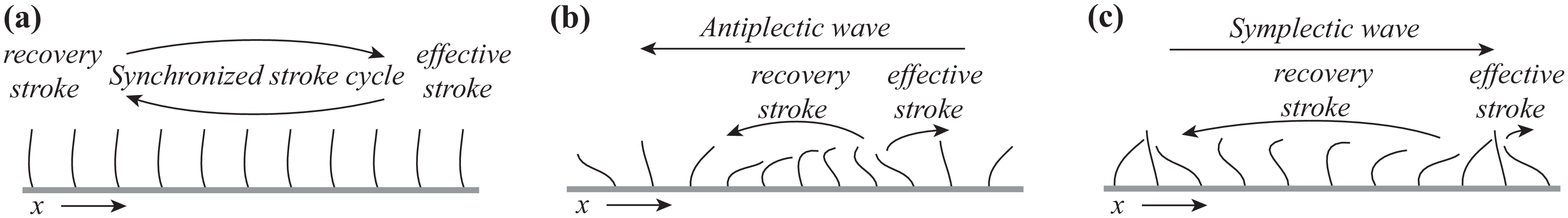}}
  	\caption[]{Schematic of ciliary array in (a) synchronized beating, (b) antiplectic wave and (c) symplectic wave. In the synchronized beating case, the phase difference between neighboring cilia is $\Delta\phi=0$; in the antiplectic wave, $0<\Delta\phi<\pi$; in the symplectic wave, $-\pi<\Delta\phi<0$. }
	\label{fig:metachronal}
\end{figure}

Motile cilia have an intricate internal structure that consists typically of a central pair of microtubules surrounded by nine microtubule doublets~\cite{vincensini2011}.  Adjacent microtubule doublets slide relative to each other under the action of ATP-fueled protein motors (dynein), generating internal moments along the cilium. These internal moments cause the cilium to deform, thus moving the surrounding fluid.  Cilia-driven flows are characterized by small Reynolds numbers, of the order $10^{-4}$ to $10^{-2}$, even in water \cite{blake1974}. At this scale, viscous forces dominate over inertial forces and Stokes equations are applicable. In order to break the time reversibility inherent in the Stokes regime, motile cilia typically use two mechanisms: (1) an asymmetric beating pattern at the level of the individual cilium (Fig.~\ref{fig:beatingpattern}), and (2) a metachronal wave pattern in cilia beating collectively (Fig.~\ref{fig:metachronal}). The asymmetric beating pattern consists of two phases: an effective stroke during which the cilium is aligned almost straight in the normal direction to the cell surface and generates a flow in the same direction as its motion, and a recovery stroke during which the cilium bends parallel to the surface and generates a weaker backward flow as it returns to its original configuration. Cilia from different cell types are known to exhibit qualitatively different kinematics \cite{blake1974}, however the physical and biological constraints that dictate these kinematics are not well understood.

The metachronal wave in collectively beating cilia is the result of all cilia performing similar beating patterns, but deforming in time with a small phase difference with respect to their neighbors (Fig.~\ref{fig:metachronal}). The spatial distribution of the cilia resulting from these phase differences leads to symmetry-breaking at the collective level and the formation of a propagating wave pattern. The mechanism causing these metachronal waves is still debated, but several recent studies have suggested that hydrodynamic interactions between neighboring cilia play a role in this coordination \cite{gueron1997, vilfan2006, lenz2006, guirao2007, niedermayer2008}.
However, hydrodynamic interactions seem to generate only one type of metachronal waves, namely, antiplectic waves that  propagate in the direction opposite to the effective stroke, see Fig.~\ref{fig:metachronal}.  Qualitatively distinct metachronal waves, such as symplectic waves that propagate in the direction of the effective stroke or waves that propagate obliquely to the effective stroke, have been experimentally observed. The reasons for this diversity in  biological systems remains largely unknown, but symplectic waves have been associated with fluid environments characterized by higher viscosity or viscoelastic behavior~\cite{knight-jones1954}. This association is based on the observation that, in symplectic waves, cilia undergoing effective stroke cluster together (Fig.~\ref{fig:metachronal}) and thus may produce larger forces to overcome the higher fluid resistance.

In this paper, we examine whether cilia beating patterns are hydrodynamically optimal by conducting a comparative study of the performance of two cilia beating patterns taken from two experimental systems: beating pattern (A) of rabbit tracheal cilia~\cite{sanderson1981,fulford1986},  and beating pattern (B) of cilia from the protozoan \textit{Opalina} \cite{sleigh1968}, see Fig.~\ref{fig:beatingpattern}. The rabbit tracheal cilia are used for fluid transport and beat in an antiplectic wave pattern whereas the \textit{Opalina}  cilia are used to propel the micro-organism and they beat in a symplectic wave pattern. We compare these two beating patterns for a range of symplectic and antiplectic metachronal waves in the context of two model systems: model (I) for swimming of a ciliated circular micro-organism, and model (II) for fluid transport/pumping by a ciliated flat surface, as depicted in Fig.~\ref{fig:models}. We employ three hydrodynamic measures of ciliary performance, including (i) average swimming speed and average pumping flow-rate, (ii) maximum forces and moments {experienced} by the cilia, (iii) swimming and pumping efficiencies. If cilia performance were hydrodynamically optimal, one would expect the tracheal cilia to outperform the \textit{Opalina} cilia in fluid transport whereas the latter would outperform the former in swimming, in one or all of these measures. This thinking proved to be too simplistic. Our results show that the performance of the \textit{Opalina} cilia is superior to that of the tracheal cilia in all the hydrodynamic-based performance measures. These findings suggest that the hydrodynamics of a single function may not explain the wide variety of cilia beating patterns observed in biological systems and that other biological parameters and constraints may be at play.


\section{Problem Formulation}
\label{sec:model}

\begin{figure}[t]
\begin{center}
{\includegraphics[scale=.5]{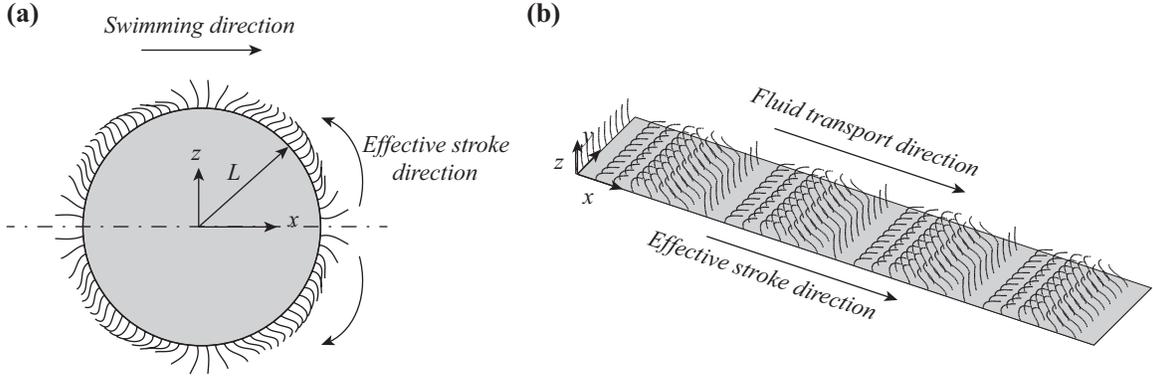}}
\end{center}
  	\caption[]{(a) Ciliated cylindrical swimmer of radius $L=1$ and $\Delta\phi=\pi/20$ (b) Ciliated carpet with double-periodic boundary in $x$ and $y$ directions and $\Delta\phi=\pi/7$. The cilia spacing in the $x$- and $y$-directions are given by $0.144L$ and $0.04L$, respectively.}
	\label{fig:models}
\end{figure}

We consider two model systems:  (I) the swimming of a ciliated cylinder and (II) the fluid pumping by a ciliated carpet, as shown in Fig.~\ref{fig:models}. The ciliated carpet model is described in details in our previous work~\cite{ding2014}. Here, we focus on describing the ciliated cylinder model and we briefly highlight the technical differences between the two models.

The cylindrical micro-swimmer of radius $L$ is covered by a finite number $N$ of discrete cilia of length $\ell$. The cilia are distributed uniformly along the cylinder's surface such that  the arc-length between the roots of two neighboring cilia is $\Delta s = 2\pi L/(N+2)$.
The cilia and their strokes are arranged symmetrically about the  $x$-axis, thus restricting the swimming motion to a pure time-varying translation in the $x$-direction,  see Fig.~\ref{fig:models}(a).

The kinematics of the beating pattern of the individual cilium can be described in a Cartesian frame attached at the base of the cilium by $\boldsymbol{\xi}_c(s,t)$, where $s$ is the arclength along the cilium's centreline from its base $(0<s<\ell)$ and $t$ is time $( 0<t< T)$.
 Two distinct cilia kinematics are depicted in Fig.~\ref{fig:beatingpattern} based on experimental data: (A) rabbit tracheal cilia and (B) cilia from the protozoan \textit{Opalina}. In both cases, the kinematics $\boldsymbol{\xi}_c(s,t)=(x_c,y_c)$ are approximated by Fourier series expansion in {$t$} and Taylor series in {$s$} with coefficients chosen to match the experimental data. Readers are referred to Fulford \& Blake~\cite{fulford1986} for details of the coefficients for the kinematics of lung cilia. We applied the same technique to the kinematics data of the \textit{Opalina}  cilia~\cite{sleigh1968}.

When each cilium goes through exactly the same cyclic beating pattern as its neighbor but with a phase difference $\Delta \phi$, a metachronal wave is generated. The frequency of the metachronal wave is equal to the frequency of the effective and recovery beat cycle. The phase difference $\Delta \phi$ determines the appearance and direction of the metachronal wave. In the synchronized case, i.e., $\Delta \phi=0$, no metachronal wave is generated. A negative phase difference amounts to compression of the cilia during the effective stroke phase and results in a symplectic wave that propagates in the same direction as the effective stroke. A positive phase difference amounts to cilia compression during the recovery stroke and results in an antiplectic wave that propagates in the opposite direction to the effective stroke, see Fig.~\ref{fig:metachronal}. The snapshot depicted in Fig.~\ref{fig:models}(a) corresponds to ciliary pattern (A)  beating in an antiplectic metachronal wave with $\Delta \phi = \pi/20$. In this remainder of this study, we use the radius $L$ of the cylinder to scale length and the inverse of the wave's angular frequency $1/\omega$ to scale time.


At zero Reynolds number, the fluid motion is governed by the non-dimensional Stokes equations and the incompressibility condition
\begin{equation}	\label{stokesequation}
\begin{aligned}
-\nabla p  & + \mu \nabla^2 \mathbf{u} =0, \qquad
\nabla \cdot \mathbf{u} =0,
\end{aligned}
\end{equation}
where $p$ is the pressure field, $\mathbf{u}$ is the fluid velocity field, $\mu$ is the dimensionless fluid viscosity. The boundary conditions in the swimmer model are
given by
\begin{equation}	\label{bc}
\begin{split}
\mathbf{u}|_{\mathrm{boundary}}&= \left\{ \begin{array}{l l}
	\mathbf{u}_c+{U}\mathbf{e}_x \qquad \text{at the cilia} \\
	{U}\mathbf{e}_x \ \qquad \qquad\text{at the cylinder}
	\end{array} \right. , 
\end{split}
\end{equation}
together with proper decay at infinity. Here, $\mathbf{u}_c$ is the prescribed velocity of the cilia  $\partial \boldsymbol{\xi}_c/\partial t$ expressed in the body frame $(\mathbf{e}_x,\mathbf{e}_z)$ attached at the center of the cylinder, and ${U}$ is the resulting (unknown) swimming speed in the $x$-direction. To determine the swimming speed $U$, recall that the total force $\mathbf{F} = F_x \mathbf{e}_x + F_z \mathbf{e}_z$ exerted by the fluid on the swimmer must be zero, since inertia is negligible. Due to the symmetry in the swimmer, $F_z$ is identically zero and does not provide an additional equation. The $x$-component of the force $F_x$ is a linear function of the swimming speed $U$ and, therefore, $F_x = 0$ provides an equation to be solved for the unknown swimming speed $U$.

Equations~(\ref{stokesequation}) and (\ref{bc}) are solved numerically using the regularized Stokelets method~\cite{cortez2001}.
This amounts to approximating each cilium by a distribution of regularized Stokeslets along
its centerline. Regularized Stokeslets are also distributed along the cylinder.
The velocity and pressure at an arbitrary point $\mathbf{x}$ in the fluid domain is approximated by the sum
over all Stokeslets ($i=1,\ldots,n_s,$ where $n_s$ is the total number of Stokeslets)
\begin{equation} \label{eq:sol}
\mathbf{u}(\mathbf{x})= \sum_{i=1}^{n_s} \mathbf{G}_s(\mathbf{x}-\mathbf{x}_i) \cdot \mathbf{f}_i, \qquad
p(\mathbf{x})=  \sum_{i=1}^{n_s} \mathbf{H}_s(\mathbf{x}-\mathbf{x}_i) \cdot\mathbf{f}_i .
\end{equation}
Here, $ \mathbf{G}_s(\mathbf{x}-\mathbf{x}_i) \cdot \mathbf{f}_i$  and $\mathbf{H}_s(\mathbf{x}-\mathbf{x}_i) \cdot\mathbf{f}_i$ represent the
velocity and pressure at $\mathbf{x}$ induced by a regularized Stokeslet of strength $\mathbf{f}_i$ located at $\mathbf{x}_i$.
Expressions for the regularized Green's tensor-valued function $\mathbf{G}_s$ and for the vector-valued function $\mathbf{H}_s$
are given by Cortez~\cite{cortez2001}. The Stokeslet strengths $\mathbf{f}_i$ depend on their
position on the cilia/cylinder and time. The values of $\mathbf{f}_i$ and $U$ are computed simultaneously by solving a set of equations arising from the no-slip boundary conditions~\eqref{bc} on the cylinder/cilia and the force balance equation $F_x=\sum_i \mathbf{f}_{i} \cdot \mathbf{e}_x=0$.

\begin{figure}[t]
\centerline{\includegraphics[width=0.35\textwidth]{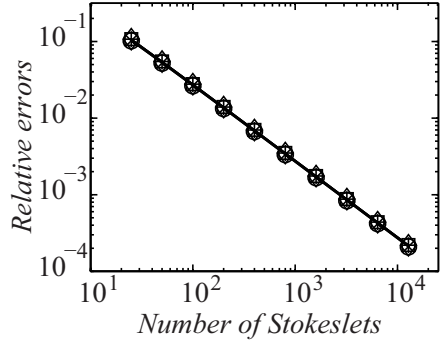}} 
  	\caption[]{The relative errors of velocities of Blake's undulating cylinder. $+, \circ, \times, \Box, \Diamond$ represent results of example 1 to 5 respectively.}
	\label{fig:blakesolution2}
\end{figure}

A note on the Stokes paradox is in order here. Stokes paradox states that for a 2D cylinder of arbitrary cross-section towed through a fluid at zero Reynolds number, there is no solution to the Stokes equations that satisfies the no-slip boundary condition at the cylinder and has finite velocity at infinity. However, the paradox does not hold for freely swimming cylinders since the total force on the swimming cylinder is zero; see, e.g., {Lauga \& Powers}~\cite{lauga2009} and references therein. This result was first proven by Blake in the case of a  cylindrical swimming of circular cross-section due to surface undulations~\cite{blake1971}.
By way of validation, we applied the framework presented here to the undulating circular cylinder~\cite{blake1971}.
The surface undulations represent the envelope of the cilia tips. Blake assumed small undulations (of order~$\epsilon$), projected them onto the surface of the cylinder,
and obtained explicit expressions (up to second order approximation in $\epsilon$) for the cylinder's swimming velocity and efficiency. He listed
numerical values for the swimming speeds and efficiencies for five examples of surface
undulations. Using the regularized Stokeslet method, we numerically solved for the swimming motion associated
with these five examples. 
{The relative errors in the average swimming velocity between the numerical solution and Blake's analytical solution are shown  in Fig~\ref{fig:blakesolution2} as functions of the number of Stokeslets on the cylinder. Clearly, the  numerical solution converges to Blake's solution as the number of Stokeslets increases.}

\begin{figure}[t]
\centerline{\includegraphics[width=0.9\textwidth]{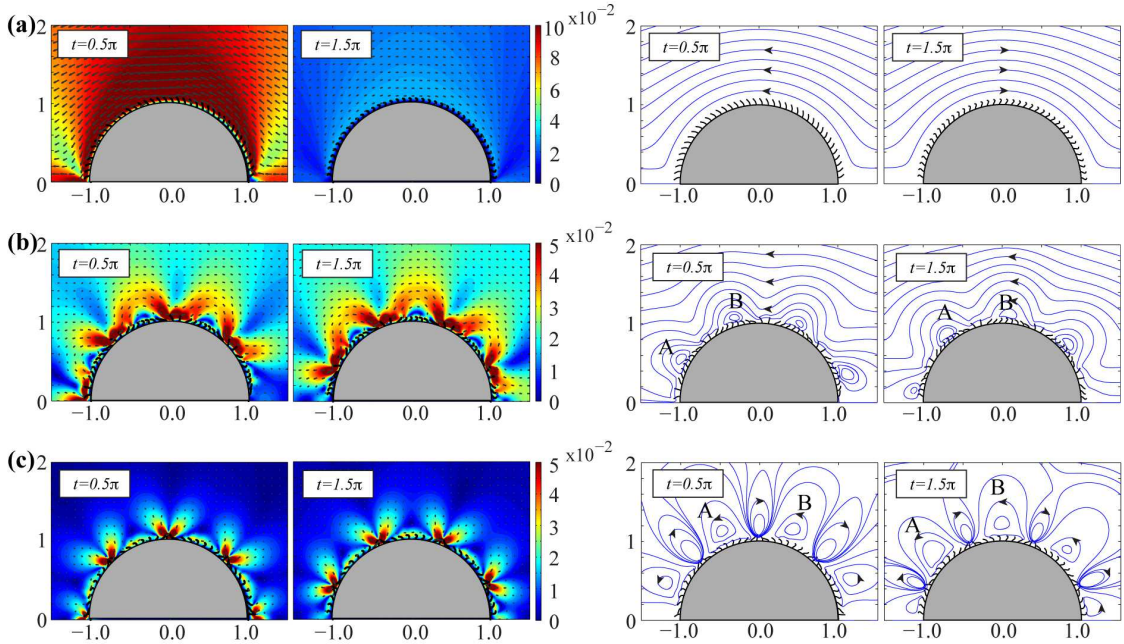}} 
  	\caption[]{Fluid velocity field and streamlines in body frame of the swimmer model with beating pattern (A). Snapshots are taken at $t=0.5\pi$ and $1.5\pi$
	for 	(a) synchronized  beating $\Delta\phi=0$, (b) antiplectic wave $\Delta\phi=0.2\pi$, and (c) symplectic wave $\Delta\phi=-0.2\pi$. Total number of cilia is $N=80$.} 
	\label{fig:contourpanel}
\end{figure}

For the fluid pumping model shown in Fig~\ref{fig:models}(b), the cilia are rooted at the plane $z=0$ with doubly-periodic boundaries in the $x$ and $y$ directions. The cilia beating motion and metachronal waves take place in the $(x,z)$-plane such that the overall fluid pumping occurs along the $x$-axis. In the $y$-direction, the cilia are more densely packed and beat synchronously. To compute the flow fields generated by such ciliary carpets, we use the regularized Stokeslet method but with three major differences from the case of the self-propelled cylinder. First, we employ the regularized Stokeslet solution~\cite{cortez2001} (i.e., the expressions for $\mathbf{G}^s$ and $\mathbf{H}^s$) associated with a three-dimensional fluid domain as opposed to those associated with a two-dimensional fluid domain in the cylinder case. Second, instead of imposing the no-slip conditions at  the bounding plane {$z=0$} using a distribution of wall-bound Stokeslets, we use the image system of Blake~\cite{blake1971b}, or, more precisely, the image system associated with a regularized Stokeslet~\cite{ainley2008}. Finally,  the solution in~\eqref{eq:sol} involves doubly-infinite sums associated with the doubly-infinite cilia domain in the $x$- and $y$-directions. In general, the fluid velocity induced by these infinite sums does not decay to zero at infinity ($z \to \infty$), but rather conditionally converges to a constant velocity  whose value is not known a priori and depends on the cilia beating pattern. In this work, we approximate those infinite sums using truncated expressions while numerically verifying that the truncated sums converge for large distances  away from the ciliated wall (for $z$ large)~\cite{ding2014}.

The net flow $Q$ transported by the ciliated carpet in the $x$-direction is equal to the flow across the $(y,z)$ half-plane, which is equal, by virtue of incompressibility, to the flow rate through any parallel half-plane, 
leading to a flow
rate per cilium $Q= \frac{1}{\mu\pi}  \int_0^\ell (\mathbf{f}_i\cdot \mathbf{e}_x) z ds$, where $\mathbf{f}_i\cdot \mathbf{e}_x$ is the Stokeslet force along the cilium in the flow direction
\cite{smith2008, ding2014}. The average flow rate per cycle is given by $\langle Q \rangle = \frac{1}{T} \int_0^T Q dt$.

To compute the internal forces and moments along each cilium, we consider each cilium to be an inextensible elastic filament~\cite{eloy2012}.
The balance of forces and moments on each cilium are given by the Kirchhoff equations for a rod, expressed in a Cartesian frame attached at
the base of that cilium,
\begin{equation}
\dfrac{\partial \mathbf{N}}{\partial s} - \mathbf{f} =0, \qquad \dfrac{\partial \mathbf{M}}{\partial s} + \mathbf{t} \! \times \! \mathbf{N}+ \mathbf{q}=0 ,
\label{eq:rod}
\end{equation}
where $ \mathbf{N}(s,t)$ and $\mathbf{M}(s,t)$ are the internal tension and elastic moment respectively,  $-\mathbf{f}(s,t)$ is the drag force
(the opposite to the force $\mathbf{f}$ exerted by the cilium on the surrounding fluid), $\mathbf{t}= \partial \boldsymbol{\xi}_c/\partial s$ is the unit
tangent to the cilium,
and $\mathbf{q}(s,t)$ is the internally generated moment per unit length.
The internal moment $\mathbf{q}$ models the discrete distribution of moments generated by the cilium dynein arms.

We evaluate the force distribution $\mathbf{f}$ along each cilium by dividing the local Stokeslet strength $\mathbf{f}_i$ by the distance between neighboring Stokeslets along that cilium. By construction, $\mathbf{f}$ accounts for the hydrodynamic coupling between all the cilia in the system (the ciliated cylinder or the ciliated carpet). This is in contrast to the single cilium case studied by Eloy and Lauga~\cite{eloy2012}.
The  elastic moment $\mathbf{M}$ along each cilium is  related to the deformation of that cilium using a linear constitutive relation
$\mathbf{M} = B \, \mathbf{D}$, where $B$ is the bending rigidity and $\mathbf{D} = \mathbf{t} \times  \mathbf{t}'$  is the Darboux vector, with the prime superscript in $\mathbf{t}'$ denoting derivative with respect to $s$.
Upon substituting into~\eqref{eq:rod}, one gets that the internal moments generated along each cilium are given by
\begin{equation}
\mathbf{q} = B \, \mathbf{t}'' \! \times \! \mathbf{t} + \mathbf{t} \! \times \!\! \int_s^\ell \mathbf{f}(\eta,t){\rm d}\eta .
\label{eq:torque}
\end{equation}

The average power $\langle P \rangle$ expended internally by the cilia to achieve the swimming or pumping task is equal to the power consumed by the internal moments $\mathbf{q}$, where only the positive works are accounted for
\begin{equation}
\langle P \rangle =\ \dfrac{1}{n}\sum_{j=1}^n  \left( \langle \int_0^\ell \max({0,\mathbf{q}\cdot\mathbf{\Omega}}) ds\rangle \right)_j ,
\end{equation}
where $\mathbf{\Omega} = \| \mathbf{\dot{t}}(s)\|  \dfrac{\mathbf{t}\times\dot{\mathbf{t}}}{\|\mathbf{t}\times\dot{\mathbf{t}}\|} $ is the angular velocity vector.
The notation $\langle \cdot \rangle$ represents average in time, whereas the outer sum corresponds to an average over cilia, $j=1,\ldots, n$. In the case of the ciliated cylinder, the power is averaged over all cilia $n = N$, whereas in the case of the ciliated carpet the average is taken over all cilia within one wavelength of the metachronal wave.
Note that accounting for only positive work means
that the dynein arms in the cilium cannot harvest energy from the fluid environment and implies that
the mean power spent by the internal moments is larger
than the power given to the fluid\cite{eloy2012}.


We present two dimensionless efficiencies: $\eta_s$ for swimming and $\eta_p$ for pumping,
\begin{equation}
\eta_s=\mu L \dfrac{\langle{U}\rangle^2}{\langle P \rangle}, \qquad  \eta_p=\mu \ell^{-3}\frac{\langle{Q}\rangle^2}{\langle{P}\rangle},
\label{eq:eff}
\end{equation}
where $\langle{U}\rangle = \frac{1}{T}\int_0^T U dt$ the average translational velocity of the swimmer.  A similar swimming efficiency has been proposed by Avron {\em et al}~~\cite{avron2004}. The pumping efficiency is consistent with that of Osterman \& Vilfan~\cite{osterman2011} and Eloy \& Lauga~\cite{eloy2012}. 

We conclude this section by presenting the parameter values we use in \S~\ref{sec:results}. The typical body length of a ciliated protozoa is of the order of $100\mu \text{m}$ and the cilium length  is about $10\mu \text{m}$~\cite{brennen1977}. Thus, we set the dimensionless  length parameters to $L = 1$ and $\ell = 0.1$. In the case of the ciliated cylinder, the spacing between the cilia is dictated by the total number of cilia on the cylinder. In the ciliated carpet case, we fix the spacing in the $x$-direction to be $0.144L$ and in the $y$-direction to be $0.04L$. Given the sparse information about the  angular frequency and bending rigidity of cilia, we use the information available for the \textit{Paramecium} as a proxy to obtain the right order of magnitude. The angular frequency in the  \textit{Paramecium}  case is about $200~ \text{rad}\cdot \text{s}^{-1}$ and the bending rigidity is estimated to be $B = 25\text{pN} \cdot\mu\text{m}^2$ \cite{eloy2012, hines1983}.
Using the characteristic length $L_c=100\mu \text{m}$, time $T_c= 1/200= 0.005\text{s}$,  cilium cross-sectional diameter $a_c = 0.12\mu \text{m}$, and viscosity $\mu_c=10^{-3}\text{Pa}\cdot\text{s}$,
the non-dimensional bending rigidity is given by $B/(\mu_c L_c^4 T_c^{-1}) \approx 1.25\times 10^{-6}$ in 3D and $B/(\mu_c a_c L_c^3 T_c^{-1}) \approx 1.04\times 10^{-7}$ in 2D. Note that using higher viscosity than that of water would result in a smaller value of the non-dimensional bending rigidity, thus reducing the relative important of the first term  on the right hand side of \eqref{eq:torque}.

{The discretization parameters are chosen as follows. The swimming cylinder is discretized using a total of $2000$ uniformly-distributed Stokeslets, which amounts to a separation distance of $3.14\times10^{-3}$ between two neighboring Stokeslets. The cilia are also uniformly-discretized using the same separation distance. Decreasing the separation distance in half (i.e., doubling the number of Stokeslets) yields results  that differ by only $1\%$ from the results reported here, thus the discretization scheme is considered to have numerically converged. In the ciliated carpet model, cilia are similarly uniformly-discretized using a separation distance of $5\times10^{-3}$ between the two neighboring Stokeslets. The doubly-infinite ciliary field is approximated using 18 copies of the computational domain (9 on each side) in each direction. More details on the convergence properties of this computation can be found in Ding \textit{et al.} 2014.~\cite{ding2014}}

\section{Results}
\label{sec:results}

The fluid velocity fields generated by the ciliated cylinder with beating pattern (A) are shown in Fig.~\ref{fig:contourpanel} for three types of metachronal coordination: (a) all the cilia beat in a synchronized way ($\Delta\phi=0$); (b) the collective ciliary beat generates an antiplectic metachronal wave ($\Delta\phi=0.2\pi$); (c) the collective ciliary beat generates a symplectic metachronal wave ($\Delta\phi=-0.2\pi$).
When the cilia beat in synchrony, they generate a relatively large flow field during their effective stroke but the flow is reversed during the recovery stroke. When the cilia beat in metachrony, the magnitude of the fluid velocity field at any time is relatively smaller than that produced by the synchronized cilia but the flow field is characterized by vortex-like structures generated by the metrachronal wave. These vortices move tangentially to the ciliated surface in the same direction as  the metachronal wave. Qualitatively similar flow fields, including the formation of vortices propagating in the same direction as the metachronal waves, are obtained  for beating pattern (B); results are not shown for brevity. Similar flow fields are also observed in the ciliated carpet model, including vortex-like structures that appear when the cilia beat in metachrony. These vortex-like structures are reminiscent to those observed in the case of Taylor's swimming sheet\cite{taylor1951, lauga2009}, albeit on one side only. In Taylor swimming sheet, the average swimming velocity scales {inversely proportional to} the wavelength. Here, the wavelength is equal to the distance between two neighboring cilia divided by the phase difference $\Delta \phi$ and times $2\pi$~~\cite{ding2014}. For convenience, we examine how the performance of the swimming and pumping systems depends on the phase difference $\Delta \phi$.


\begin{figure}[t]
\centerline{\includegraphics[width=\textwidth]{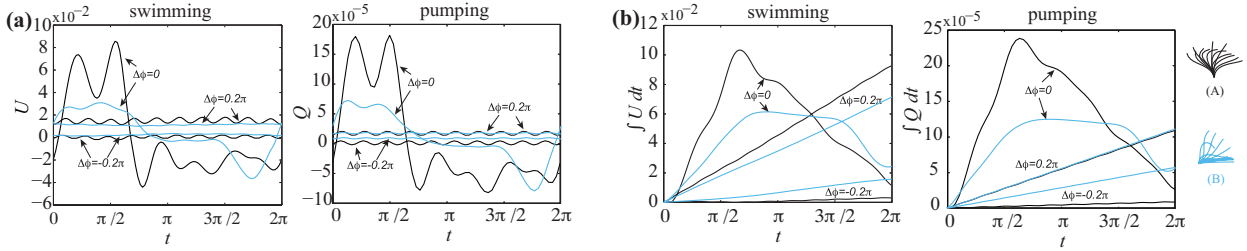}}
  	\caption[]{Swimming velocities and flow rates per cilium as functions of time for $\Delta\phi=0, 0.2\pi, -0.2\pi$ and for patterns (A) black and (B) blue.
	}
	\label{fig:typicalvel}
\end{figure}

 Fig.~\ref{fig:typicalvel} compares the swimming velocity $U$ (ciliated cylinder) and  pumping flow rate $Q$ (ciliated carpet) for the three cases $\Delta \phi = 0, 0.2 \pi,$ and $-0.2\pi$ depicted in Figs.~\ref{fig:contourpanel}. In each case and  in the context of each model, we compare cilia patterns (A) and (B).
 Clearly, the maximum values of $U$ and $Q$ occur for $\Delta \phi = 0$ (Fig.~\ref{fig:typicalvel}(a)), however the average speed $\langle U\rangle$ and  average flow rate $\langle Q\rangle$  are substantially increased when $\Delta \phi = 0.2\pi$ and somewhat decreased when $\Delta \phi = - 0.2\pi$ (Fig.~\ref{fig:typicalvel}(b)),  suggesting the existence of optimal metachronal waves that maximize swimming and pumping.

\begin{figure}[t]
\begin{center}
{\includegraphics[width=0.6\textwidth]{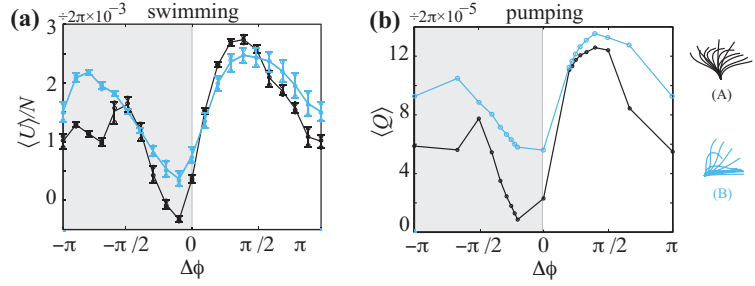}}
\end{center}
  	\caption[]{Average swimming velocity (normalized by the number of cilia) and average flow rate per cycle as a function of  $\Delta \phi$ for patterns (A) black and (B) blue. Symplectic waves (shaded in grey) are characterized by negative $\Delta \phi$ while antiplectic waves have positive $\Delta \phi$. For the ciliated cylinder in (a), the solid lines correspond to the average values of three different cases $N=50,60$, and $70$. The variation in the average swimming velocity between these cases is indicated by the error bars at each value of $\Delta \phi$.
	}
	\label{fig:disp}
\end{figure}

We examine the dependence of the average swimming speed (ciliated cylinder) and average flow rate (ciliated carpet) {over one cycle} on all phase differences $\Delta \phi$ from $-\pi$ to $\pi$, for the two cilia patterns (A) and (B), see Fig.~\ref{fig:disp}. We emphasize that pattern (A) corresponds to pumping-specialized cilia  that beat antiplectically in their natural system (rabbit trachea) whereas  pattern (B) corresponds to swimming-specialized cilia  that beat symplectically in their natural system (the protozoan \textit{Opalina}). Therefore, we expected to find an optimal antiplectic wave ($\Delta \phi >0$)  for which pattern (A) maximizes pumping, and an optimal symplectic wave ($\Delta \phi <0$) for which pattern (B) maximizes swimming. Instead, we found  that the optimal performance of patterns (A) and (B) occurs for antiplectic metachronal waves, with pattern (A) slightly outperforming pattern (B) in swimming, and pattern (B) slightly outperforming pattern (A) in pumping  (Fig.~\ref{fig:disp}).
Surprisingly, the \textit{Opalina}-ciliary pattern (B) outperforms the tracheal-ciliary pattern (A) in pumping for all values of $\Delta \phi$.

\begin{figure}[t]
\centerline{\includegraphics[width=\textwidth]{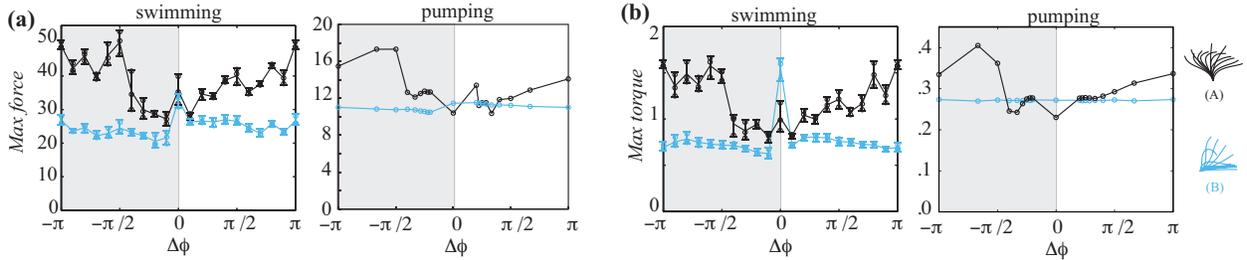}}
  	\caption[]{Maximum forces and moments experienced by the individual cilia as a function of $\Delta \phi$ for patterns (A) black and (B) blue for the same parameter values as those in {\em Fig.~\ref{fig:disp}}.
  	}
	\label{fig:maxF}
\end{figure}

We next examine the performance of the two ciliary patterns (A) and (B) in terms of the  maximum forces and moments the cilia experience as a function of $\Delta \phi$.
The maximum forces and moments a cilium can generate are limited by the cilium biological structure. It's been postulated that when the hydrodynamic load on the cilia is larger than the forces and moments that the individual cilium can generate, e.g., in fluid environments characterized by higher viscosity, cilia tend to cluster together during the effective stroke to collectively produce the forces and moments needed to overcome the higher fluid resistance, without overloading the individual cilium and causing it to collapse. This argument explains the emergence of the symplectic metachronal wave as an adaptation to large hydrodynamic load on the cilia. However, little is known about the effect of the maximum moments afforded by the cilium's internal structure in shaping the beating pattern of the individual cilium in response to the hydrodynamic loading.  This effect is to be distinguished from efficiency considerations, which we discuss next. Besides efficiency, the maximum moments generated by the individual cilium could have an important selective effect on the cilia beating pattern. We postulate that the fitness of the beating pattern can be assessed by the maximum moments experienced at the cilium level that are required to produce a desired net displacement or a desired net flow at the system level.

Fig.~\ref{fig:maxF} depicts the maximum hydrodynamic forces and moments experienced by patterns (A) and (B) for all values of $\Delta \phi$. In the ciliated cylinder model, we used a representative cilium located midway between the two ends of the ciliated cylinder, away from the cilia near the symmetry axis $\mathbf{e}_x$. Clearly,
the forces and moments experienced by pattern (B) are smaller than or equal to those experienced by pattern (A) for all valued of  $\Delta \phi$, except for $\Delta \phi=0$. As far as we know, the case $\Delta \phi=0$ has never been reported in natural ciliated systems. 
This suggests that pattern (B) is better suited for environments that impose large hydrodynamic load on the cilia.

\begin{figure}[t]
\centerline{\includegraphics[width=\textwidth]{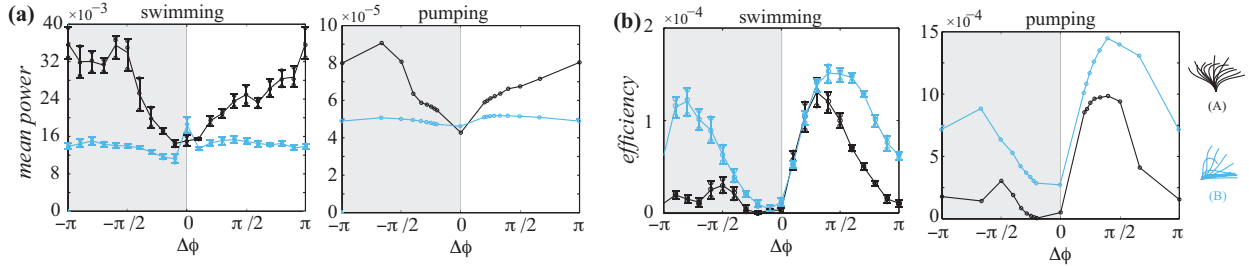}}
  	\caption[]{(a) Average power and  (b) swimming and pumping efficiencies as a function of $\Delta \phi$ for patterns (A) black and (B) blue for the same parameter values as those in {\em Figs.~\ref{fig:disp}} and {\em~\ref{fig:maxF}}.
  	}
	\label{fig:power}
\end{figure}

We conclude this section by comparing the average power and efficiency of patterns (A) and (B)  for all values of $\Delta \phi$.
It is clear from Fig.~\ref{fig:power}(a) that beating pattern (B) consumes less energy than beating pattern (A) in both the swimmer and pumping systems, except when the cilia are beating synchronously. The power expended by beating pattern (A) is more sensitive to the phase difference $\Delta \phi$ than that of beating pattern (B).
The swimming efficiencies of the beating pattern (A) and (B) are almost the same in the range of $0<\Delta\phi<0.3\pi$. Beating pattern (B) shows advantages over (A) in rest of the region. In the pumping model, beating pattern (B) outperformed (A) at every phase difference, with maximum efficiency reaching 1.5 times that of (A), see Fig.~\ref{fig:power}(b).

\section{Discussion}
\label{sec:conc}
We examined the hydrodynamic performance of two cilia beating patterns, reconstructed from experimental data. These two beating patterns were chosen as representative examples of pumping-specialized cilia (pattern (A)) and swimming-specialized cilia (pattern (B)). In their respective natural systems, they correspond to antiplectically-coordinated cilia (pattern (A)) and  symplectically-coordinated cilia (pattern (B)).
{To our best knowledge, the exact parameters of these metachronal waves, such as phase difference, have neither been quantified nor associated with a particular function. However, our results indicate that two major functions of cilia activity, swimming and pumping, are highly sensitive to the phase difference of the metachronal wave. To allow for comparison to the biological system, we therefore} attempted to compute the phase difference between neighboring cilia from still images\cite{sanderson1981, tamm1970}. These computations lack accuracy due to missing information about the orientation of the ciliated surface relative to the field of vision of the camera and about the wave direction but they indicate that  the antiplectic tracheal cilia, pattern (A), beat at $\Delta \phi \approx 0.02\pi$ whereas $\Delta \phi \approx -0.03\pi$ for the symplectic \textit{Opalina} cilia, pattern (B).

We compared the hydrodynamic performance of {the} two cilia beating patterns as a function of the metachronal coordination (phase difference $\Delta \phi$) in the context of (I) the swimming of a ciliated cylinder and (II) the fluid pumping by a ciliated carpet.  We used three performance measures: (i) average swimming speed and average pumping flow rate; (ii) maximum internal moments generated along the cilia; and (iii) swimming and pumping efficiencies based on the ratio of the average swimming/flow rate to the average power expenditure. Measures (i) and  {(iii)} have been already proposed and employed in computing optimal cilia beating patterns, both individually~\cite{eloy2012} and collectively\cite{osterman2011}. However, little attention has been paid to measure  {(ii)}, the absolute value of the internal moments generated along the cilium, as having an important selective effect on the cilium's beating pattern. This proposition is justified on the grounds that the internally-generated moments are limited by the cilium's biological structure. Our results can be summarized as follows:
\vspace{-0.5\baselineskip}
\begin{enumerate} \addtolength{\itemsep}{-0.5\baselineskip}
\item In both the swimming and pumping models, according to measures (i) swimming speed/pumping rate and (iii) swimming/pumping efficiency, one can identify two optimal metachronal coordination for each beating pattern (see Figs.~\ref{fig:disp} and~\ref{fig:power}): one antiplectic  and one symplectic. This is consistent with previous computational results~\cite{khaderi2011,khaderi2012, ding2014}. The optimal values of $\Delta \phi$ are summarized in Table~\ref{tab:optphase}. Interestingly, these optimal values do not  match  the values of $\Delta \phi$ that we computed from the still images of the cilia. This discrepancy suggests that the metachronal coordination in these natural systems is not hydrodynamically optimal. 
\item According to measure (ii)  maximum internal moments, a clear optimal coordination cannot be identified (see Fig.~\ref{fig:maxF}).  The maximum internal moments generated by pattern (B), which naturally inhabits a very viscous environment,
seem to vary smoothly as a function of $\Delta \phi$ whereas the dependence of the maximum internal moments on $\Delta \phi$ in pattern (A) is irregular. This observation suggests that pattern (B) is more resilient to changes in the cilia metachronal coordination than pattern (A), in the sense that it requires smaller adjustment in it internal moments.
\item  Pattern (B), taken from swimming-specialized cilia that beat symplectically in their natural system, outperforms pattern (A), taken from pumping-specialized cilia that beat antiplectically in their natural system,  in almost all three performance measures. In particular, we found that pattern (B)  is  more efficient than pattern (A) for all metachronal waves. These results are surprising because they challenge the notion that hydrodynamic efficiency dictates the cilia beating pattern.
\vspace{-0.5\baselineskip}
\end{enumerate}
 \begin{table}
  \caption{ \label{tab:optphase} \footnotesize The optimal phase difference $\Delta \phi$ corresponding to  measures (i) top row and (iii) bottom row.}
 \begin{adjustwidth}{-2in}{-2in}
 \centering
 {\footnotesize
 \begin{tabular}{c |c c| c c|}
 \multicolumn{1}{c}{}&\multicolumn{4}{c}{\textbf{Beating Pattern (A)}}  \\
 \cline{2-5}
 \multicolumn{1}{c}{}&\multicolumn{2}{|c|}{swimming} & \multicolumn{2}{|c|}{pumping} \\
 \cline{2-5}
  & antiplectic  & symplectic & antiplectic  & symplectic  \\
 \cline{2-5}
& $0.4\pi$ & $-0.5\pi$ & $0.4\pi$ &$-0.5\pi$ \\
& $0.3\pi$ & $-0.5\pi$ &$0.4\pi$ &$-0.5\pi$ \\
 \cline{2-5}
 \end{tabular}
 \hspace{0.1in}
  \begin{tabular}{c |c c| c c|}
 \multicolumn{1}{c}{}&\multicolumn{4}{c}{\textbf{Beating Pattern (B)}}  \\
 \cline{2-5}
 \multicolumn{1}{c}{}&\multicolumn{2}{|c|}{swimming} & \multicolumn{2}{|c|}{pumping} \\
 \cline{2-5}
  & antiplectic  & symplectic & antiplectic  & symplectic  \\
 \cline{2-5}
&$0.4\pi$ &$-0.8\pi$ &$0.4\pi$ &$-0.67\pi$\\
&$0.4\pi$ &$-0.8\pi$ &$0.4\pi$ &$-0.67\pi$ \\
 \cline{2-5}
 \end{tabular}
 }
 \end{adjustwidth}
 \vspace{-1.25\baselineskip}
 \end{table}

Our findings suggest that factors other than the ones accounted for in our study could be at play in determining the cilia beating kinematics. For example, cilia often function in viscoelastic fluid environments\cite{sleigh1989}, which induce effects not captured by the Newtonian model employed here. The effect of viscoelasticity on the swimming speed of micro-organisms has caused some controversy as to whether it enhances or hinders  swimming, with recent reports suggesting that both are possible depending on the swimming stroke and the viscoelastic properties of the medium \cite{lauga2007, spagnolie2013, shen2011}. In the cilia-related literature, a few models, such as the {viscoelastic traction layer model}\cite{lubkin2007}, have been proposed to account for the effect of a viscoelastic layer at the cilia tips but, to our knowledge, the optimal beating of a cilium in a viscoelastic fluid remains an open problem.

Motile cilia are known to serve not only in moving the surrounding fluid for swimming or pumping functions but also in sensing of environmental cues \cite{bloodgood2010}.  The way by which ciliary motion serves these two tasks simultaneously is not very well understood. Experimental\cite{supatto2008,shields2010}and computational\cite{ding2014} studies seem to imply that chemical diffusion is not  the sole mechanism for sensing and that cilia-generated flows enhance diffusion by chaotic advection and mixing.  This suggests that  effective sensing may be another measure for evaluating the cilia beating kinematics. In many aquatic species, cilia also serve to generate feeding currents\cite{pepper2013,riisgaard2010} which require a delicate balance between bringing sufficient nutrients from the far field and slowing them down for consumption in the ciliary near field. Again, the effect of such multi-tasking on the individual beating kinematics is not well understood.

Other effects that we have neglected in our study include {three-dimensional beating kinematics. This is because we estimated that the cilium bending motion is mostly planar in the case of the two beating patterns examined in this study, thus justifying a planar beating model. In particular, we quantified the out-of-plane component of cilia motion by comparing the projected cilium length in the image sequence of the beat cycle to absolute cilium length as reported in literature\cite{sanderson1981,sleigh1968}. We found little change in the cilium's overall length over one beating cycle ({with standard deviation about $3\%$ and $5\%$ for beating pattern (A)\cite{sanderson1981} and (B)\cite{sleigh1960}, respectively}), indicating that the cilium bending motion is mostly two-dimensional. Out-of-plane cilia kinematics and other three-dimensional effects, such as obliquely propagating metachronal waves, may play important roles in other systems, however, and are easy to implement in future extensions of this study.}

Finally, we note that, although the cilium's internal structure of microtubules and dynein motors has been highly conserved throughout evolution and across cilia from various cell types, other structural and physiological constraints at the cell surface could influence the beating kinematics, especially in light of recent reports suggesting that {spatiotemporal coordination of ciliary beating strongly dependent on cytoskeletal connections between neighboring cilia\cite{werner2011, vladar2012}}.




\bibliography{references}

\end{document}